\documentclass[a4paper,12pt]{article} 

\usepackage{epsfig}
\usepackage{amssymb}
\usepackage{amsmath}
\usepackage{amsfonts}
\usepackage{hyperref}
\usepackage{cite}

\def\Ca{{C^{}_A}}
\def\Tr{{T^{}_R}}
\def\Cf{{C^{}_F}}
\def\Nc{N_c}

\newcommand{\eps}{\epsilon}
\newcommand{\ep}{\epsilon}
\newcommand{\be}{\begin{equation}}
\newcommand{\ee}{\end{equation}}

\newcommand{\ORD}[1]{\mathop{}\!\mathcal{O}\left(#1\right)}
\newcommand{\GF}[1]{\mathop{}\!\Gamma\left(#1\right)}
\newcommand{\GFP}[2]{\mathop{}\!\Gamma^#1\left(#2\right)}
\newcommand{\GENHYPGF}[5]{F_{#1 #2}\left ( #3,#4,#5\right )}
\newcommand{\HYPGF}[4]{\GENHYPGF{2}{1}{#1,#2}{#3}{#4}}
\newcommand{\GPL}[2]{\text{G}_{#1}(#2)}

\newcommand{\Li}{{\rm Li}}

\def \mySij {\widetilde{S_{ij}}}
\def \myIij {\widetilde{\mathcal I_{ij}}}

\def \Tcol {{\bf T}}
\def \intSS {\mathcal S{\hspace{-5pt}}\mathcal S}

\def \lnss {\ln(s^2)}
\def \lncs {\ln(c^2)}
\def \lnt  {\ln2}
\def \lnopss {\ln\left(1+s^2\right)}
\newcommand {\lnsk}[1] {\ln^{#1}(s^2)}
\newcommand {\lnck}[1] {\ln^{#1}(c^2)}
\newcommand {\lntk}[1]  {\ln^{#1}2}
\newcommand {\lnopsk}[1]  {\ln^{#1}\left(1+s^2\right)}

\begin{document}

\vspace{-5.0cm}
\begin{flushright}
IPPP/18/48, TTP18-020
\end{flushright}

\vspace{2.0cm}

\begin{center}
{\Large \bf
The double-soft integral for an arbitrary angle between hard radiators
}\\
\end{center}

\vspace{0.5cm}

\begin{center}
Fabrizio Caola$^{1}$, Maximilian Delto$^{2}$, Hjalte Frellesvig$^{2}$, Kirill Melnikov$^{2}$.\\
\vspace{.3cm}
{\it
{}$^1$IPPP, Durham University, Durham, UK\\
{}$^2$Institute for Theoretical Particle Physics, KIT, Karlsruhe, Germany
}

\vspace{1.3cm}

{\bf \large Abstract}
\end{center}
\noindent
We consider the double-soft limit of a generic QCD process involving
massless partons and integrate analytically the double-soft eikonal
functions over the phase-space of soft partons (gluons or quarks)
allowing for an arbitrary relative angle between the three-momenta
of two hard massless radiators. This result provides one of the
missing ingredients for a fully analytic formulation of the nested
soft-collinear subtraction scheme described in Ref.~\cite{Caola:2017dug}. 

\thispagestyle{empty}

\pagenumbering{arabic}

\allowdisplaybreaks
                    
\section{Introduction}

A precise description of hard processes offers an exciting opportunity
to discover or constrain physics beyond the Standard Model at the LHC
using indirect methods. Such a
description is  based on the collinear factorization framework that
emphasizes the  importance of understanding 
partonic cross sections in higher orders of perturbative
QCD.  Currently, it is possible to compute most processes
of phenomenological interest at a fully-differential
level at leading and next-to-leading orders in perturbative
QCD, and $2 \to 1$ and $ 2 \to 2 $ processes
at  next-to-next-to-leading order (NNLO). 

An important recent development in the field of precision
collider physics is 
the start 
of ``mass production'' of NNLO QCD results for major $2 \to 2$ LHC
processes such as  $pp \to t \bar t$~\cite{Czakon:2013goa,Czakon:2015owf},
$pp \to 2j$~\cite{Currie:2016bfm,Currie:2017eqf}, 
$pp \to V+j$~\cite{Boughezal:2015dva,Boughezal:2015ded,Ridder:2015dxa,Gehrmann-DeRidder:2017mvr,Campbell:2016lzl},
$pp \to H+j$~\cite{njet1,Boughezal:2015dra,Chen:2014gva,Chen:2016zka},
the $t$-channel single top production~\cite{Brucherseifer:2014ama,Berger:2016oht},
Higgs production in weak boson
fusion \cite{Cacciari:2015jma,Cruz-Martinez:2018rod} etc. General purpose public numerical
codes also became available recently~\cite{mcfm_colsing,matrix}. This progress
happened because a number of computational schemes, both of the 
slicing  type and the subtraction type
\cite{ant,Czakon:2010td,Czakon:2011ve,Czakon:2014oma,
  Boughezal:2011jf,Cacciari:2015jma,qt1,qt2,njet1,njet2,colorful,Caola:2017dug},
have matured enough to be used in complex realistic  calculations.

Nevertheless, in spite of these successes, it is fair
to say that none of the suggested schemes are fully  optimal. 
This is unfortunate as it
can  limit our ability to make precise 
predictions for higher multiplicity processes at the LHC in the future. 
Hence, further developments of subtraction methods
are  welcome.  Motivated by these considerations, two
of us in collaboration with R.~R\"ontsch have recently
proposed \cite{Caola:2017dug} a modification
of the subtraction scheme described in
Refs.~\cite{Czakon:2010td,Czakon:2011ve,Czakon:2014oma}. 
A key element in our proposal is the double-soft limit
defined as follows.

We consider the double-real emission contribution to NNLO QCD corrections
to the production of an arbitrary final state $X$ in hadron collisions. 
Specifically, we are interested in  the $X + f$ final state, where $f$ are either
two gluons or a quark-antiquark pair. We assign four-momenta 
$k_{4,5}$ to the two additional partons, and consider the 
double soft configuration $k_{4,5} \to 0$, with no particular hierarchy
between $k_4$ and $k_5$. 
It is well known that soft emissions factorize.
Indeed, 
in the soft approximation parton emission does not
change the kinematics of the final state $X$ and does not affect
infra-red safe observables. Moreover, 
the matrix element squared of the process $ ij \to X+f$
factorizes into  a color-correlated emissionless matrix element
squared for the process $ij \to X$ 
and a universal  eikonal function that depends on momenta of
hard radiators that are present in either the initial or the final state,  and  
the momenta $k_4$ and $k_5$ of the soft partons.  

An important ingredient of any NNLO subtraction scheme is the
integral of the double-soft eikonal function over the phase-space 
of the two extra partons $f$,  subject to kinematic constraints.  
In the  framework of  Ref.~\cite{Caola:2017dug},
the following constraints on energies of the soft partons are  imposed
\be
k_4^0 < E_{\rm max},\;\;\;\; k_5^0 < k_4^0.
\label{eq4}
\ee
In Ref.~\cite{Caola:2017dug}, double-soft  integrals with constraints as in Eq.~(\ref{eq4})
were computed  numerically for the case when hard emittors are back-to-back. 
Although this is adequate for the color-singlet production processes considered 
in~\cite{Caola:2017dug}, in more complicated cases 
the numerical approach becomes cumbersome, since it requires 
a non-trivial continuation of phase-space integrals beyond four space-time 
dimensions (see~\cite{Czakon:2010td,Czakon:2011ve,Boughezal:2013uia,Czakon:2014oma} for details).
Moreover, beyond the back-to-back limit, the double-soft integrals become functions
of an angle between the three-momenta of the hard radiators. Since these angles change from event to
event, the required numerical computations become quite expensive. 

In what follows, we show how to  overcome these issues and   present 
an analytic computation of  
the double-soft integrals required for the  description of
NNLO real emission contributions 
to an arbitrary process.
The rest of this paper
is organized as follows. In Section~\ref{sec:dsoft} we introduce
our notation, present  relevant formulas for the 
double-soft limit and  define the integrals that need to be computed.  
In Section~\ref{sec:int} we discuss  how to use 
differential equations to find the phase-space integrals. 
In Sec.~\ref{sec:bc} we explain how to fix the boundary conditions 
needed to fully
reconstruct the required integrals from the
differential equations. In Section~\ref{sec:finres} we present our final results for the
integrals of the double-soft eikonal functions. 
We conclude in Section~\ref{sec:conclusions}.

\section{The double-soft current and its integration}
\label{sec:dsoft}
In this section, we consider the double-soft limit of a generic scattering process. 
It is well known that soft emissions factorize. We now recall  basic features
of this factorization, following closely Ref.~\cite{Catani:1999ss}. Interested readers should
consult Ref.~\cite{Catani:1999ss} for further details. 

In QCD, soft emissions involve  non-trivial color correlations. It is then convenient to 
introduce a color basis $|c_1,...,c_n\rangle$, and write a generic scattering amplitude as
\be
\mathcal M^{c_1,...,c_n}(p_1,...,p_n) = \left\langle c_1,...,c_n|\mathcal M(p_1,...,p_n)\right\rangle, 
\ee
where $c_i$ are the color indices. It is also useful to  associate a color charge $\Tcol_i$ 
with the emission of soft gluons off a parton $i$. Its action is defined as 
\be
\begin{gathered}
\left\langle c_1,...,c_i,...,c_m,a|\Tcol_{i}|b_1,...,b_i,...b_m\right\rangle
=\delta_{c_1 b_1}...T^{a}_{c_i b_i}...\delta_{c_m b_m},
\end{gathered}
\ee
where $a$ is the gluon color index ($a=1,...,\Nc^2-1)$ and $T^a_{c_i b_i} = i f_{a c_i b_i}$ if
parton $i$ is a gluon, $T^a_{c_i b_i} = t^a_{c_i b_i}$ if $i$ is a quark,  and 
$T^a_{c_i b_i} = \bar t^a_{c_i b_i} = - t^a_{b_i c_i}$ if $i$ is an antiquark. Here $f_{abc}$ 
and $t^{c}_{ab}$ are
the generators of the
${\rm SU}(N_c)$  Lie algebra in the adjoint and fundamental representations, respectively. 
The color charge operators satisfy
\be
T_i^a T_j^a = \Tcol_i \cdot \Tcol_j = \Tcol_j\cdot \Tcol_i,~~~~
\Tcol_i^2 = C_i,
\ee
with $C_i = \Ca = \Nc $ if $i$ is a gluon and $C_i = \Cf = (\Nc^2-1)/(2\Nc)$ if $i$ is a
quark or an  antiquark. Note also that each vector $|\mathcal M(p_1,...,p_n)\rangle$ is a color-singlet
state, which implies
\be
\sum_{i=1}^{n} \Tcol_i |\mathcal M(p_1,...,p_n)\rangle = 0. 
\ee

Using this notation, the matrix element squared for the process $ij\to X +f(k_4,k_5)$ 
in the double-soft limit $k_4,k_5\to 0$ can be written as~\cite{Catani:1999ss}
\be
\begin{split} 
& |{\cal M}(g_4,g_5;\{p\})|^2
\approx \big[ g_{s,b}^2 \mu^{2 \ep}\big]^2
\left [ \frac{1}{2} \sum \limits_{ij,kl}^{n}
  S_{ij}(k_4) S_{kl}(k_5)
   |{\cal M}^{(i,j)(k,l)}(\{p \})|^2
\right.  \\
&  \left . 
~~~~~~~~~
- C_A \sum \limits_{i < j}^{n} \mySij(k_4,k_5)
|{\cal M}^{(ij)}(\{p\})|^2
\right ],
\label{eq:dsgg} 
\end{split}
    \ee
if $f(k_4,k_5)$ are two gluons, and
\be
\begin{split}
& |{\cal M}(q_4,\bar q_5;\{p\})|^2
\approx \big[g_{s,b}^2 \mu^{2\ep}\big]^2
\Tr \sum_{i<j} \myIij(k_4,k_5)|{\cal M}^{(ij)}(\{p\})|^2,
\end{split}
\label{eq:dsqq}
\ee
if $f(k_4,k_5)$ are  a quark and an antiquark.
In Eqs.~(\ref{eq:dsgg},\ref{eq:dsqq}),
 ``$\approx$'' means that we
only consider the most singular contribution in the double-soft limit, and
\be
\begin{gathered}
|{\cal M}^{(ij)(kl)}(\{p \})|^2 \equiv \left\langle
\mathcal M(p_1,...,p_n)| \left\{\Tcol_i\cdot \Tcol_j,
\Tcol_k\cdot\Tcol_l\right\} 
| \mathcal M(p_1,...,p_n)\right\rangle,
\\
|{\cal M}^{(ij)}(\{p \})|^2 \equiv \left\langle
\mathcal M(p_1,...,p_n)| \Tcol_i\cdot \Tcol_j | \mathcal M(p_1,...,p_n)\right\rangle.
\end{gathered}
\ee
We also used $g_{s,b}$ to denote  the bare QCD coupling constant and $\Tr=1/2$. 
The sums in Eqs.~(\ref{eq:dsgg},\ref{eq:dsqq}) run  over
all pairs of hard color-charged radiators.
The functions $\mySij$ and $\myIij$ read
\be
\begin{split}
&\mySij(k_4,k_5) = 2 S_{ij}(k_4,k_5) - S_{ii}(k_4,k_5)- S_{jj}(k_4,k_5),
\\
&\myIij(k_4,k_5) = 2 \mathcal I_{ij}(k_4,k_5) - \mathcal I_{ii}(k_4,k_5)- \mathcal I_{jj}(k_4,k_5).
\end{split}
\ee
All  other eikonal functions are defined as follows \cite{Catani:1999ss} 
\be
\begin{split}
  & S_{ij}(k) = \frac{p_i \cdot p_j}{(p_i \cdot k) (p_j \cdot k) },
  \\
  & S_{ij}(k_4,k_5) = S_{ij}^{\rm so}(k_4,k_5)
  -\frac{2 p_i \cdot p_j}{k_4 \cdot k_5 \big[p_i \cdot (k_4+k_5)\big]\big[  p_j \cdot (k_4+k_5)\big] }
  \\
  &
    + \frac{( p_i \cdot k_4) (p_j \cdot k_5) + (p_i \cdot k_5)(p_j \cdot k_4)
  }{\big[p_i \cdot(k_4+k_5)\big] \big[p_j \cdot (k_4+k_5)\big] }
  \left [
    \frac{(1-\ep)}{(k_4 \cdot k_5)^2 }
    -\frac{1}{2} S_{ij}^{\rm so}(k_4,k_5) \right ],
\end{split} 
\ee
where $\ep = (4-d)$/2 with $d$ being the space-time dimensionality,
\be
\begin{split} 
& S_{ij}^{\rm so}(k_4,k_5) = \frac{p_i \cdot p_j}{k_4 \cdot k_5}
\left ( \frac{1}{(p_i \cdot k_4) (p_j \cdot k_5) }
+ \frac{1}{(p_i \cdot k_5) (p_j \cdot k_4) } \right )
\\
& - \frac{(p_i\cdot p_j)^2}{(p_i\cdot k_4) (p_j \cdot k_4)
(p_i \cdot k_5) (p_j \cdot k_5) },
\end{split} 
\ee
and
\be
\mathcal I_{ij} = \frac{(p_i\cdot k_4)(p_j\cdot k_5)+(p_i\cdot k_5)(p_j\cdot k_4)
-(p_i\cdot p_j)(k_4\cdot k_5)}{(k_4\cdot k_5)^2 \big[p_i\cdot(k_{4}+k_5)\big]
\big[p_j\cdot(k_{4}+k_5)\big]}.
\ee

According to the computational scheme described in Ref.~\cite{Caola:2017dug},  
    the double-soft matrix elements in Eqs.~(\ref{eq:dsgg},\ref{eq:dsqq})
    should be integrated over the three-momenta of soft partons,
    subject  to the constraints in Eq.~(\ref{eq4}).  
    It follows that a double-soft contribution
    to any  differential cross section can be constructed
    if the following integrals are known
\be
\begin{split}
  &     \intSS_{ij,kl}  = \int [d k_4] [d k_5]
  \theta(E_{\rm max} - k_4^{0}) \theta( k_4^{0} - k_5^{0})
  S_{ij}(k_4) \; S_{kl}(k_5),
\\
&     \intSS^{(gg)}_{ij}  = \int [d k_4] [d k_5]
  \theta(E_{\rm max} - k_4^{0}) \theta( k_4^{0} - k_5^{0})
\mySij(k_4,k_5),
\\
& \intSS^{(q \bar{q})}_{ij} = -2 \int [d k_4] [d k_5]
  \theta(E_{\rm max} - k_4^{0}) \theta( k_4^{0} - k_5^{0})
\myIij(k_4,k_5),
\label{eq:dsint}
\end{split}
\ee
where the factor -2 in $\intSS^{(q\bar{q})}_{ij}$ is introduced for convenience. 
In Eq.~(\ref{eq:dsint}) we introduced the short-hand notation
\be
[dk_i] = \frac{{\rm d}^{(d-1)}k_i}{2 k_i^0(2\pi)^{d-1} }.
\ee

As explained in Ref.~\cite{Caola:2017dug}, the energy ordering $E_4>E_5$ 
accounts for the $1/2!$ symmetry factor relevant for $gg$ emission. We find
it convenient to use the same phase-space parametrization for $q\bar q$
emission as well. Since $\myIij(k_4,k_5)=\myIij(k_5,k_4)$, the full result
in the $q\bar q$ case is twice the result that is obtained by
imposing the $E_4>E_5$ ordering. 

We satisfy constraints in Eqs.~(\ref{eq4},\ref{eq:dsint}) 
    by choosing the following parametrization of the energies of the two soft partons
    \be
    k_4^{0} = E_{\rm max}\; \xi, \;\;\; k_5^{0} = E_{\rm max}\; \xi z,
    \;\;\; 0 < \xi< 1,\;\;\;  0 < z < 1. 
    \label{eq12new}
    \ee
We note that  integrals in Eq.~\eqref{eq:dsint} depend on the relative  angles between the three-momenta
of the hard partons but not on their energies. 
The first integral $\intSS_{ij,kl}$ in Eq.~\eqref{eq:dsint} is easy to compute since
it is  a product
of two single-gluon eikonal functions.  For further convenience, we   introduce the
following notation
\be
p_i = E_i (1,\vec n_i),\;\;
k_4 = E_{\rm max} \; \xi \; (1,\vec n_4),\;\;
k_5 = E_{\rm max} \; \xi z \; (1,\vec n_5),\;\;
\label{eq16a}
\ee
where $\vec n_i^2 = 1$. Moreover, we will use 
\be
\rho_{ij} = 1 - \vec n_i \cdot \vec n_j.
\ee
in what follows.

To compute $\intSS_{ij,kl}$,
we integrate over $\xi$ and $z$ and obtain
\be
\intSS_{ij,kl} = \frac{E_{\rm max}^{-4\ep}}{8 \ep^2}
\int \frac{{\rm d} \Omega_4}{2 (2\pi)^{d-1}} \frac{\rho_{ij}}{\rho_{i4} \rho_{j4}}
\int \frac{{\rm d} \Omega_5}{2 (2\pi)^{d-1}} \frac{\rho_{kl}}{\rho_{k5} \rho_{l5}}.
\ee
The angular integrals over the emission angles of the gluons $g_4$ and
$g_5$ completely factorize and can be easily performed.
One obtains \cite{vanNeerven:1985xr}
\be
\int \frac{{\rm d} \Omega_4}{2(2\pi)^{d-1}} \frac{\rho_{ij}}{\rho_{i4} \rho_{j4} }
= -\frac{2}{\epsilon} \left [ \frac{1}{8 \pi^2} \frac{(4\pi)^\ep}{\Gamma(1-\ep)}
  \right ] (2 \rho_{ij} )^{-\ep} {\cal F}_{ij}, 
\ee
where
\be
\begin{split}
{\cal F}_{ij} =& \left [ \frac{\Gamma(1-\ep)^2}{\Gamma(1-2\ep)} \right ]
\left ( \frac{\rho_{ij}}{2} \right )^{1+\ep}
~F_{21}\left (1,1,1-\ep,1-\frac{\rho_{ij}}{2} \right )
\\
&
= 1 + 
\ep^2\left[\Li_2\left(\frac{\rho_{ij}}{2}\right)- \zeta_2\right] + 
{\cal O}(\ep^3).
\end{split}
\ee
Finally, we obtain
\be
\intSS_{ij,kl} = \frac{(2E_{\rm max})^{-4\ep}}{2 \ep^4}
\left [ \frac{1}{8 \pi^2} \frac{(4\pi)^\ep}{\Gamma(1-\ep)}
  \right ]^2 \left(\frac{\rho_{ij}}{2}\right)^{-\ep}
\left(\frac{\rho_{kl}}{2}\right)^{-\ep}  {\cal F}_{ij} {\cal F}_{kl}.
\label{eq:ss_uncorr}
\ee
It is straightforward to obtain
the expansion of the hypergeometric function through ${\cal O}(\ep^4)$
using existing computer algebra packages \cite{Maitre:2005uu}.

The non-trivial part of the computation requires the calculation of
the correlated emission terms
$\intSS_{ij}^{(gg/q\bar q)}$ as a function of the scattering angle between the two
hard partons.  We describe such a calculation in the next section.

\section{Double soft integrals}
\label{sec:int}
In this section, we describe the calculation of the correlated contributions 
to the eikonal integrals    $\intSS_{ij}^{(gg/q\bar q)}$ defined in
Eq.~\eqref{eq:dsint}. For definiteness, we focus on the $\intSS_{ij}^{(gg)}$ computation. 
The calculation of  $\intSS_{ij}^{(q \bar q)}$  proceeds in a similar fashion. 

We use the fact that $\mySij(k_4,k_5)$ is a  homogeneous function
  of the soft momenta $k_{4,5}$ and of the hard momenta $p_{i,j}$. This implies that if we use the
  parametrization of the four-momenta as in Eq.~\eqref{eq16a}, we can
  integrate over the variable $\xi$.
 After the $\xi$  integration,  we separate the integration over
  $z$ and write
  \be
  \intSS^{(gg)}_{ij} =
  -\frac{(2E_{\rm max})^{-4\ep}}{\ep}
  \int \limits_{0}^{1} {\rm d}z \; G_{ij}(z).
  \label{eq10}
  \ee
The function $G_{ij}(z)$ is defined as 
  \be
  G_{ij} = 
  \int [d k_4] [d k_5] \mySij( k_4,k_5) \delta( 2 P \cdot k_4 - 1)
   \delta(2 P \cdot k_5 -z),
\label{eq11}
   \ee
  and the four-momentum $P$ is a  time-like vector
  $P=(1,\vec 0)$. Although the final result for double
  soft integrals does not depend on the normalization
  of the four-momenta of hard radiators, 
  when  computing individual contributions 
  we will use $p_{i,j} = 1/2(1,\vec n_{i,j})$. 
  The two $\delta$-functions in Eq.~(\ref{eq11}) provide constraints
  on the energies of the two gluons $k_{4,5}$; their arguments are chosen to make
  them ``propagator-like'' for reasons that will become clear later.

  To calculate $G_{ij}(z)$ we need to integrate $\mySij(k_4,k_5)$ 
  over the phase-space of the two gluons with energy constraints
  shown in Eq.~\eqref{eq11}.
   We do this by mapping these phase-space
  integrals onto loop integrals following Ref.~\cite{Anastasiou:2002yz}.
  After defining integral  families, we apply the integration-by-part
  identities to reduce the number of independent integrals that need
  to be computed and to derive differential equations that these integrals
  satisfy.

  We identify 19 master integrals to be calculated. To display them, we
  introduce seven propagator-like  structures 
\begin{align}
\begin{aligned}
&   D_1 =   2p_1 \cdot k_4, \;\; D_2 =  2p_2 \cdot k_4, \;\; D_3 = 2p_1 \cdot k_5,\;\;
   D_4 =2p_2 \cdot k_5, \\
& D_5= p_1 \cdot \left( k_4 +  k_5 \right),\;\;
  D_6= p_2 \cdot \left( k_4 +  k_5 \right),\;\;
  D_7=2 k_4 \cdot k_5,
\label{eq24new}
\end{aligned}
\end{align}
and define 
\be
\left \langle \frac1{\prod_iD_i^{\alpha_i}} \right \rangle = 
\int \frac{[d k_4] [d k_5] \delta( 2 P \cdot k_4 - 1)
   \delta(2 P \cdot k_5 -z) }{\prod_iD_i^{\alpha_i}}.
\ee
With this notation, we require the following integrals\footnote{In the
$q\bar q$ case, only the integrals $I_{1-3}$ and $I_{12-14}$ contribute.}
\begin{align}
\begin{aligned}
  & I_1  = \left \langle 1 \right \rangle,\;  I_2 = \left \langle \frac1{D_5} \right \rangle,\;
  I_3  = \left \langle \frac1{D_5^2} \right \rangle,\;I_4   = \left \langle \frac1{D_4D_5} \right \rangle,\;
  \\
  & 
  I_5  = \left \langle \frac1{D_4D_5^2} \right \rangle,\;
  I_6  = \left \langle \frac1{D_4D_5D_7} \right \rangle,\;
  I_7    =  \left \langle \frac{D_1}{D_4D_5D_7} \right \rangle,\;
  \\
  & 
  I_8   =  \left \langle \frac1{D_1D_6} \right \rangle,\;
  I_9    =  \left \langle \frac1{D_1D_6^2} \right \rangle,\;
  I_{10}  =  \left \langle \frac1{D_1D_6D_7} \right \rangle,\;
  \\
  & 
  I_{11} =  \left \langle \frac{D_4}{D_1D_6D_7} \right \rangle,\;
  I_{12}  =  \left \langle \frac1{D_5D_6} \right \rangle,\;
  I_{13} =  \left \langle \frac1{D_5^2D_6} \right \rangle,\;
  \\
  &
  I_{14} =  \left \langle \frac{D_4}{D_5^2D_6} \right \rangle,\;
  I_{15} =  \left \langle \frac1{D_1D_5D_6} \right \rangle,\;
  I_{16} =  \left \langle \frac1{D_4D_5D_6} \right \rangle,\;
\\
&
  I_{17} =  \left \langle \frac1{D_1D_4D_5D_6} \right \rangle,\; 
  I_{18} = \left \langle \frac1{D_1D_4D_7} \right \rangle,\;
I_{19} = \left \langle \frac1{D_1^2D_4D_7} \right \rangle.
\end{aligned}
\label{eq:listofint}
\end{align}
  These master   integrals are functions of the energy fraction $z$ and of the
  relative angle $\theta$ between the two hard radiators $i$ and $j$. 
  To compute them, we use differential equations.

  In principle, we can write differential equations for master integrals
    in both $z$ and $\theta$. As it is easy to see
  from their definition, the two integrals $I_{18}$ and $I_{19}$ are homogeneous
  in $z$; this implies that the $z$-differential equation does not give any
  non-trivial information in this case. 
    Therefore, we computed these 
  two integrals by solving   the differential
  equation with respect to the scattering angle. The boundary conditions
  for these differential equations were  determined
  from the values of $I_{18,19}$  computed in a situation when
  the three-momenta of the radiators
  are back-to-back, i.e. $\theta = \pi$.   We find the following results 
\allowdisplaybreaks
\begin{align}
  \label{eq18aa}
& I_{18} = \frac{N_\ep}{x z^{1+2\ep}} \times \bigg\{ \frac{3}{\ep^2} - \frac1{\ep} \big[ 12 + 6 \GPL{0}{x} \big] + \big[ 12 + \pi^2 + 24 \GPL{0}{x} \notag
\\
&\hphantom{I_{18} + \ep \big[}
+12 \GPL{0,0}{x} - 8 \GPL{1,0}{x} \big] \notag\\
& \hphantom{I_{18}} + \ep \big[ -4 \pi^2 -24 \GPL{0}{x} -2 \pi^2 \GPL{0}{x} + \frac{4\pi^2}{3} \GPL{1}{x} -48 \GPL{0,0}{x}  \notag\\
& \hphantom{I_{18} + \ep \big[} 
+ 32 \GPL{1,0}{x}
- 24 \GPL{0,0,0}{x} + 16 \GPL{0,1,0}{x} + 16 \GPL{1,0,0}{x} \notag
\notag \\
& \hphantom{I_{18} + \ep \big[} 
- 8 \GPL{1,1,0}{x} -18 \zeta_3 \big] \notag \\
& \hphantom{I_{18}} + \ep^2 \big[ 4 \pi^2 - \frac{\pi^4}{10} + 8 \pi^2 \GPL{0}{x} - \frac{16\pi^2}{3} \GPL{1}{x} + 48 \GPL{0,0}{x} + 4 \pi^2 \GPL{0,0}{x}\notag \\
& \hphantom{I_{18} + \ep^2 \big[} -\frac{8\pi^2}{3} \GPL{0,1}{x} -32 \GPL{1,0}{x} - \frac{8\pi^2}{3} \GPL{1,0}{x} + \frac{4\pi^2}{3} \GPL{1,1}{x} \notag \\
& \hphantom{I_{18} + \ep^2 \big[} + 96 \GPL{0,0,0}{x} - 64 \GPL{0,1,0}{x} -64 \GPL{1,0,0}{x} + 32 \GPL{1,1,0}{x} \notag \\
& \hphantom{I_{18} + \ep^2 \big[} + 48 \GPL{0,0,0,0}{x} -32 \GPL{0,0,1,0}{x} - 32 \GPL{0,1,0,0}{x} + 16 \GPL{0,1,1,0}{x} \notag \\
& \hphantom{I_{18} + \ep^2 \big[} -32 \GPL{1,0,0,0}{x} +24\GPL{1,0,1,0}{x} + 16 \GPL{1,1,0,0}{x} - 8 \GPL{1,1,1,0}{x} \notag \\
& \hphantom{I_{18} + \ep^2 \big[} +72 \zeta_3 +36 \zeta_3 \GPL{0}{x}  -24\zeta_3 \GPL{1}{x} \big] \bigg\},
\notag \\
& I_{19} = \frac{N_\ep}{x^2 z^{1+2\ep}} \times \bigg\{ \frac{3}{\ep^2} - \frac1{\ep}  \big[ 6x + \GPL{0}{x} \big] \notag \\
& \hphantom{I_{19}} + \big[ 14x + \pi^2 -30  + 4x \GPL{0}{x} +12 \GPL{0,0}{x} - 8 \GPL{1,0}{x} \big] \\
& \hphantom{I_{19}} + \ep \big[ 26 x + 18 - \frac{2\pi^2}{3}x + 60 \GPL{0}{x} -2 \pi^2 \GPL{0}{x} -12 x \GPL{0}{x} \notag \\
& \hphantom{I_{18} + \ep \big[} 
+ \frac{4\pi^2}{3} \GPL{1}{x}
-8x\GPL{0,0}{x} + 8x \GPL{1,0}{x} - 24 \GPL{0,0,0}{x} \notag \\
& \hphantom{I_{19} + \ep \big[} 
+ 16 \GPL{0,1,0}{x} + 16 \GPL{1,0,0}{x} 
- 8 \GPL{1,1,0}{x} - 18 \zeta_3 \big] \notag \\
& \hphantom{I_{19}} + \ep^2 \big[ 54 - 10 \pi^2 - \frac{\pi^4}{10} - 90 x + 2 \pi^2 x -36 \GPL{0}{x} - 4 x \GPL{0}{x} 
\notag \\
& \hphantom{I_{19} + \ep^2 \big[}
+ \frac{4\pi^2}{3} x \GPL{0}{x} 
- \frac{4\pi^2}{3} x \GPL{1}{x} 
 -120 \GPL{0,0}{x} + 4\pi^2 \GPL{0,0}{x} \notag \\
& \hphantom{I_{19} + \ep^2 \big[} 
+ 24 x \GPL{0,0}{x} 
- \frac{8\pi^2}{3} \GPL{0,1}{x} + 80 \GPL{1,0}{x} 
- \frac{8\pi^2}{3} \GPL{1,0}{x} \notag \\
& \hphantom{I_{19} + \ep^2 \big[}
- 24 x \GPL{1,0}{x} 
+ \frac{4\pi^2}{3} \GPL{1,1}{x} + 16 x \GPL{0,0,0}{x}  
-8 x \GPL{0,1,0}{x}  \notag \\
& \hphantom{I_{19} + \ep^2 \big[} 
-16 x \GPL{1,0,0}{x} 
+ 8x \GPL{1,1,0}{x} + 48 \GPL{0,0,0,0}{x} 
-32 \GPL{0,0,1,0}{x} \notag \\
& \hphantom{I_{19} + \ep^2 \big[} 
-32 \GPL{0,1,0,0}{x} 
+ 16 \GPL{0,1,1,0}{x} -32 \GPL{1,0,0,0}{x} 
+ 24 \GPL{1,0,1,0}{x} \notag \\
& \hphantom{I_{19} + \ep^2 \big[} 
+ 16 \GPL{1,1,0,0}{x} 
- 8 \GPL{1,1,1,0}{x} + 12 \zeta_3 x 
+ 36 \zeta_3 \GPL{0}{x} 
\notag \\
& \hphantom{I_{19} + \ep^2 \big[} 
-24 \zeta_3 \GPL{1}{x} \big] \bigg\}. \notag
\end{align}
In writing the expressions for $I_{18,19}$, we used the following expression
for the normalization factor 
\be
N_\epsilon = \left[\frac{\Omega^{(d-1)}}{(2\pi)^{d-1} 2^{2-2\ep}}\right]^2
= \left[\frac{1}{8\pi^2}\frac{(4\pi)^\ep}{(1-2\ep)}\frac{\Gamma(1-\ep)}{\Gamma(1-2\ep)}
\right]^2.
\label{eq:nep}
\ee
Also, $x$ is the sine squared of half the relative angle between the three-momenta
of hard radiators $x = \sin^2 \delta$, $\delta = \theta/2$,
and $\GPL{a_1,a_2,...,a_m}{x}$ are the standard
Goncharov polylogarithms.

The situation with the remaining seventeen integrals is rather different.
Indeed, many of them couple to each other and the majority of them are not homogeneous
functions of $z$.   Although it is possible  to
use  differential equations w.r.t. the relative energy and angle to determine the integrals
also in this case, 
we found it more convenient to consider the differential
  equation in the energy fraction $z$ and to determine the full dependence
  on the angle between the hard radiators
   by computing boundary conditions as functions
  of $\theta$. 
We did not use a canonical form \cite{Henn:2013pwa}  for the $z$ differential equation. 
  In fact, it is relatively straightforward to achieve a canonical
  form for the first eleven  integrals but after that it becomes
  much more difficult to do so.  
  However, we managed to re-write the system of differential equations 
  in such a way that integrating it by expanding
    master integrals order-by-order in $\epsilon$
    becomes possible.  In principle, this is absolutely sufficient for solving the system
    of differential equations. 
    A possible drawback of this approach is that intermediate results tend to be 
    quite cumbersome. This is, however, easy to deal with once all the expressions for 
    the integrals are substituted to obtain the physical result.

 The differential equations in $z$ are of the  following form
  \be
  \frac{\partial }{\partial z} \vec I(z,\delta) = \hat A(\ep,z,\delta) \vec I,
   \ee
where $\hat A(0,z,\delta)$ is a triangular matrix with vanishing diagonal elements. 
   To integrate these differential equations, it is important
   to expose the dependence of the matrix $\hat A$ on inverse
   powers of the monomials of $z$.  This dependence is characterized
   by elements of the list shown below 
  \be
  \left\{z, ~(1+z), ~(\sin^2 \delta + z),~ (1+ z\;\sin^2 \delta ),~
  \sqrt{ (1-z)^2 \sin^2 \delta + 4 z} \right\}.
\label{eq12}
\ee
The integration of the system of differential equations is greatly simplified if
its coefficients are rational functions of the integration variables. 
To achieve this, we rationalize 
the square root in Eq.~\eqref{eq12} using the 
following change of variables
\be
z = \frac{\left ( 1 - \cos \delta \; t \right )
(\cos \delta -t ) }{t \sin^2 \delta }.
\label{eq15a}
\ee
It leads to 
\be
\sqrt{ (1-z)^2 \sin^2 \delta + 4 z} = \frac{\cos \delta}{\sin \delta}
\frac{(1-t)(1+t)}{t}.
\ee
In addition to making the square root rational, the  variable transformation Eq.~(\ref{eq15a})
also maps all other $z$-dependent
monomials  in Eq.~\eqref{eq12}  onto rational functions
of $t$. We obtain
\be
\begin{split} 
& 1+z = \frac{\cos \delta}{t \sin^2\delta}
  (t - e^{i \delta}) ( t - e^{-i \delta}),
  \\
  & \sin^2 \delta +z  = \frac{\cos \delta}{ t \sin^2 \delta}
   \left ( t- a_+ \right ) \left (t - a_- \right ),
  \\
  & 1+ z\sin^2 \delta = \frac{\cos \delta}{t}
   \left (t-b_+ \right ) \left (t - b_-\right), 
\end{split} 
  \ee
  where
  \be
\begin{split}
&  a_{\pm} = \cos \delta \left ( 1 + \frac{\sin^2 \delta}{2} \right )
  \pm i \sqrt{1 - \cos^2\delta \left (1+\frac{\sin^2 \delta}{2} \right )^2 },
  \\
& b_{\pm} = \frac{\cos \delta}{2} \pm i \sqrt{1 - \frac{\cos^2 \delta}{4} }.
\label{eq18a}
\end{split}
 \ee
 As the  result of the $z \to t$ mapping, we obtain
 a system of linear differential equations for the seventeen integrals
 with rational coefficients
 \be
 \frac{\partial  }{\partial t } \vec I (t) = \hat B(\delta,t,\epsilon)
 \vec I.
 \ee
 Since the matrix $B$ is a rational function of $t$,  integration
 over $t$ can be   performed in terms of Goncharov polylogarithms
 in a straightforward manner. This gives the result up to an integration
constant that must be determined by matching to  appropriate boundary
conditions. We discuss the computation of the latter in the next section.

\section{Boundary conditions}
 \label{sec:bc}
 As explained in the previous section, we only integrate the $t$ differential
  equation, without considering a differential equation in the scattering angle $\theta$.
    We then need to compute
  the master integrals at a given value of $t$ (or $z$)  as a function
  of $\theta$.  It is natural to consider the boundary condition 
  at $z=0$, which corresponds to the situation where one of the two soft
  particles is much softer than the other. 
  Not only is this the simplest kinematic point  where such a computation
  can be performed, but it is also very useful for 
  the subsequent integration over $z$ in Eq.~\eqref{eq10} since
  that  integration is, in fact, singular at $z=0$. 

  The computation of boundary conditions is relatively straightforward for the
  majority of the master integrals but there are  a few of them that require
  some effort.  We will illustrate the relevant techniques by considering
  two representative examples.

  The simplest master integral is the phase-space itself. It can
  be computed in a straightforward way by
  first integrating over energies and then over emission angles.
  The result reads
  \be
  I_1 = \int [dk_4][dk_5] \delta(2 k_4 \cdot P - 1)
  \delta (2 k_5 \cdot P - z ) =
  N_\ep z^{1-2\ep},
  \ee
  where the normalization factor $N_\ep$ is defined in Eq.~\eqref{eq:nep}.
  A significantly more complex integral is $I_{13}$, which reads (cf. Eqs.~(\ref{eq:listofint},\ref{eq24new}))
  \be
I_{13} = \int  \frac{ [dk_4][dk_5] \delta(2 k_4 \cdot P - 1)
  \delta (2 k_5 \cdot P - z )
}{\big[p_i \cdot (k_4+k_5)\big]^2 \big[p_j \cdot (k_4+k_5)\big]}.
\ee
Upon integrating over gluon energies, we obtain
\be
I_{13} =
64 N_\ep
z^{1-2\ep} \int 
\frac{[d \Omega_4] [d \Omega_5] }{(\rho_{i4}+z \rho_{i5})^2
    (\rho_{j4} + z \rho_{j5})},
\label{eq20}
\ee
where we introduced  the normalized solid angle integration measure as
\be
[d \Omega_i] = \frac{ {\rm d} \Omega_i^{(d-1)}}{\Omega^{(d-1)}},~~~~~~
\int[d\Omega_i] = 1.
\ee
By inspecting Eq.~\eqref{eq20},
it is easy to see that,  at small $z$, the master
integral $I_{13}$ scales as  $z^{-1}$. Therefore, we need to compute it
to first subleading power to determine the integration constant. To accomplish this,   
we first combine denominators using Feynman parameters
\be
\frac{1 }{(\rho_{i4}+z \rho_{i5})^2 (\rho_{j4} + z\rho_{j5})}
= 2 \int \limits_{0}^{1} \frac{{\rm d} x \; x }{
  ( \rho_{4\eta}  + z \rho_{5\eta})^3},
\ee
where $\rho_{4,5\; \eta} = 1 - \vec n_{4,5} \cdot \vec \eta $ and
$\vec \eta = \vec n_i x + (1-x) \vec n_j$.
We then use this representation in Eq.~\eqref{eq20} and integrate over
directions of the gluon $g_4$. We obtain
\be
\int
\frac{ [d \Omega_4] }{(  1-\vec n_4 \cdot \vec \eta
  + z ( 1- \vec n_5 \cdot \vec \eta ) )^3}
= \frac{F_{21}(3,1-\epsilon,2-2\epsilon,\frac{2 \eta}{1+\eta+z\rho_{5\eta}})
  }{(1+z\rho_{5\eta} - \eta)^3}, 
\label{eq27}
\ee
with $\eta = \sqrt{\vec \eta \cdot \vec \eta} = \sqrt{1 - 4x(1-x)\sin^2\delta}$.
We still need to integrate the  right hand side of Eq.~\eqref{eq27} over $x$ and direction of the vector $\vec n_5$
to obtain $I_{13}$, i.e.
\be
\begin{split}
I_{13} = & 
64 N_\ep
z^{1-2\ep} 
\int \limits_{0}^{1} 2x {\rm d} x\;
\int 
\frac{[d \Omega_5]}{(1+\eta + z \rho_{5\eta} )^3}
\\
&
\times F_{21}
\left (
3,1-\epsilon,2-2\epsilon,\frac{2 \eta}{1+\eta+z\rho_{5\eta}}
\right ).
\end{split}
\label{eq25}
\ee
It is quite obvious that such computations simplify
dramatically if the expansion in small $z$ is possible at early stages
of the computation. Unfortunately, the hypergeometric function  in Eq.~\eqref{eq25} can not be expanded
in powers of $z$ because 
the maximal value of its argument  in the $z \to 0$ limit is one and 
the hypergeometric
function in Eq.~\eqref{eq25}  is non-analytic there. 
To transform the integrand
in Eq.~\eqref{eq25} to a suitable form, we use the  standard transformation
for hypergeometric functions that connects $F_{21}(...,y)$ with
$F_{21}(...,1-y)$. We also note that  since
$\eta$ is invariant under the replacement $x \leftrightarrow  (1-x) $, 
one can replace $2 x$ with  $1$ in 
the integrand  in Eq.~\eqref{eq25} without affecting the value of the
integral $I_{13}$.  Splitting the integral into two contributions as the consequence
of the hypergeometric transformation, we write 
\be
\begin{split}
& I_{13} = 64 N_\ep
  z^{1-2\ep} \left (
\frac{(-2+8\ep^2)}{(1+\ep)(2+\ep)}
  T_{13}^{(a)} 
+
\frac{\Gamma(2-2\ep) \Gamma(2+\ep)}{2\Gamma(1-\ep)}
T_{13}^{(b)}
\right ),
\label{eq30a}
\end{split}   
\ee
with
\be
\begin{split} 
&
T_{13}^{(a)} =\int \limits_{0}^{1}  {\rm d} x\;
\int \frac{[d \Omega_5]}{(1+\eta +\kappa )^3}
 F_{21} \left (
3,1-\epsilon,3+\epsilon,\frac{1-\eta +\kappa}{1+\eta+\kappa}
\right ),
\\
&
T_{13}^{(b)} =\int \limits_{0}^{1}  {\rm d} x\;
\int \frac{[d \Omega_5]}{(1+\eta + \kappa )^{1-\ep}(1-\eta+\kappa)^{2+\ep}}
\\
& \;\;\; \times F_{21} \left (
-1-2\ep,1-\epsilon,-1-\ep,\frac{1-\eta + \kappa}{1+\eta+\kappa}
\right ),
\end{split}
\label{eq31a}
\ee
where $\kappa = z \rho_{5\eta}$.

In principle, the hypergeometric functions in Eq.~\eqref{eq31a} can
be directly expanded in powers of $z$, since the goal
of the transformations described above has already
been achieved. However, 
the remaining integrations over $x$ and the directions
of the vector $\vec n_5$ would have been quite difficult in this case. 
Fortunately, there exists another transformation of the hypergeometric
function that reduces the complexity of the remaining integrations
dramatically. 
It reads
\be
F_{21}(a,b,a-b+1,y) = (1+y)^{-a}F_{21} \left (
\frac{a}{2},\frac{a}{2} + \frac{1}{2}, a-b+1, \frac{4y}{(1+y)^2} \right ).
\label{eq28}
\ee
As we will see, this transformation  completely removes square roots from the computation
of the boundary conditions. 

We begin by  applying this relation to the computation of $T_{13}^{(a)}$.
We note that $T_{13}^{(a)} \sim {\cal O}(1)$ in the $z \to 0$ limit,
so that it contributes directly to the subleading term in
the $z$-expansion of $I_{13}$.  
Therefore,  we are allowed to set $z =0$ in the computation
of $T_{13}^{(a)}$.
We find
\be
\lim_{z\to 0}T_{13}^{(a)} =
\frac{1}{8} \int \limits_{0}^{1} {\rm d} x \;
F_{21}\left ( \frac{3}{2},2,3+\ep,1-\eta^2 \right ).
\ee
To compute this integral, we write  the function $F_{21}$ as
hypergeometric series and integrate over $x$ using $1-\eta^2 = 4x (1-x)
\sin^2\delta$. We find a very simple result 
\be
T_{13}^{(a)} = \frac{1}{8} F_{21} \left (1,2,3+\ep,\sin^2\delta \right ) + \mathcal O(z).
\label{eq34a}
\ee

The computation of $T_{13}^{(b)}$ is somewhat more complex, primarily because
the $z \to 0$ limit can not be taken directly. Using the transformation
Eq.~\eqref{eq28}, we obtain
\be
\begin{split} 
& T_{13}^{(b)} =
\int \limits_{0}^{1} {\rm d} x
\int [{\rm d} \Omega_5]
\frac{2^{1+2\ep} (1+\kappa)^{1+2\ep}}{(1-\eta^2 + 2\kappa + \kappa^2)^{2+\ep}}
\\
& \times F_{21}\left ( -\frac{1}{2} -\ep,-\ep,-1-\ep,
\frac{1-\eta^2+2\kappa + \kappa^2}{(1+\kappa)^2}
\right ).
\label{eq32}
\end{split}
\ee
To proceed further we note that if we write the hypergeometric
function in Eq.~\eqref{eq32} as the standard hypergeometric series, we can take the
$z \to 0$ limit in all but the first two terms of the
expansion. Hence, we consider the contribution of these two terms separately. We write
\be
T_{13}^{(b)} = T_{13}^{(b),1} + T_{13}^{(b),2} + T_{13}^{(b),\Sigma}.
\label{eq36a}
\ee
We begin with  the computation of $T_{13}^{(b),\Sigma}$. We use the series
  representation of the hypergeometric function in Eq.~\eqref{eq32},
  set $z$ to zero, integrate over $x$ and $\vec n_5$  and obtain
  \be
  T_{13}^{(b),\Sigma} =
  \frac{\Gamma^2(-1-\ep) \Gamma(2-\ep)}{16 \Gamma(-2-2\ep)
    \Gamma(-\ep) }
  \left ( \sin\delta \right )^{-2\ep} \;F_{21}(1,2-\ep,3,\sin^2 \delta ) 
+\mathcal O(z).
  \ee

  Next, we compute the contribution $T_{13}^{(b),2}$ that
  arises if we take the second term in the series representation 
  of the hypergeometric function in Eq.~\eqref{eq32}.
  This term can be written as
  \be
\begin{split} 
  &   T_{13}^{(b),2} =
  -\frac{\ep(1+2\ep)}{(1+\ep)} 2^{2\ep} W^{(b),2}
  \\
  & W^{(b),2} =
  \int [{\rm d} \Omega_5]
  \int \limits_{0}^{1} {\rm d} x
  \frac{( 1+ \kappa)^{-1+2\ep} }{(  1- \eta^2 + 2\kappa + \kappa^2
  )^{1+\ep}}.
  \end{split} 
\label{eq35}
\ee
Since $1-\eta^2 = 4x(1-x) \sin^2\delta$ and $\kappa \sim {\cal O}(z)$,
it is impossible to expand the integrand in Eq.~\eqref{eq25}
in Taylor series in $z$.
Nevertheless, to obtain  an approximation to  $W^{(b),2}$ at small values of 
$z$ by expanding the integrand, we can follow 
ideas about  asymptotic expansions of Feynman diagrams  known as
the ``strategy of regions''~\cite{Beneke:1997zp}.

The integral in Eq.~\eqref{eq35} has, obviously, three regions:
{\it i}) $x \sim {\cal O}(z)$, {\it ii}) $(1-x) \sim {\cal O}(z)$ and
{\it iii}) $ x \sim (1-x) \sim 1$. The first two (soft) regions
give identical contributions;
we consider one of them and multiply the result by two.
We  refer to the third region as the ``hard region''. We therefore
write 
\be
W^{(b),2} = W^{(b),2}_H + 2 W^{(b),2}_S + \mathcal O(z).
\ee
The contribution of the hard region is obtained upon expanding the integrand 
in Eq.~\eqref{eq35}  in Taylor series 
in powers of $z$. Since we are interested in the ${\cal O}(z^0)$
term only, we obtain
\be
W^{(b),2}_H = \int [d \Omega_5] \int \limits_{0}^{1}
\frac {{\rm d} x }{\left ( 4 x (1-x) \sin^2 \delta \right )^{1+\ep}}
= (4\sin^2 \delta)^{-1-\ep} \frac{\Gamma^2(-\ep)}{\Gamma(-2\ep)}.
\label{eq37}
\ee
To compute the soft contribution, we focus on the region
$x \sim z$. We  expand  the integrand assuming $x \sim z \ll 1$ and
extend the upper $x$ integration boundary to infinity~\cite{Beneke:1997zp}.
We obtain 
\be
\begin{split}
W^{(b),2}_S & = \int [d \Omega_5]
\int \limits_{0}^{\infty}
\frac{{\rm d} x}{ \left ( 4 x \sin^2 \delta + 2z (1-\cos \theta_5)
   \right )^{1+\ep}
}
\\
& =\frac{z^{-\ep}}{2^{1+2\ep} \ep \sin^2 \delta }
\frac{\Gamma(1-2\ep)\Gamma(2-2\ep)}{\Gamma(2-3\ep)\Gamma(1-\ep)}.
\label{eq38}
\end{split}
\ee

The calculation of $T_{13}^{(b),1}$ proceeds along the same lines. The only
complication is that, since $T_{13}^{(b),1} \sim {\cal O}(z^{-1})$,
we need to expand the soft contribution in Taylor series in
$x \sim z \ll 1$ to first subleading term. Apart from additional algebraic complexity, 
this does not lead to any conceptual complications. 
We provide the results
of the calculation for completeness. We write 
\be
T^{(b),1} = T^{(b),1}_H + 2 T^{(b),1}_S + \mathcal O(z),
\ee
where
\be
T^{(b),1}_H=\frac{2^{1+2\ep}}{\left ( 4 \sin^2 \delta \right )^{2+\ep}}
\frac{\Gamma^2(-1-\ep)}{\Gamma(-2-2\ep)},
\label{eq40}
\ee
and
\be
\begin{split}
T^{(b),1}_S &= \frac{z^{-1-\ep}}{4 (1+\ep) \sin^2 \delta}
(1-z) \frac{\Gamma(-2\ep)\Gamma(2-2\ep)}{\Gamma(1-3\ep)\Gamma(1-\ep)}
\\
& 
+\frac{z^{-\ep} ( 2+3\ep(1+\ep)\sin^2\delta)}
     {4\ep(1+\ep) \sin^4 \delta}
  \frac{\Gamma(1-2\ep)\Gamma(2-2\ep)}{\Gamma(2-3\ep)\Gamma(1-\ep)}.   
\end{split} 
\label{eq41}
\ee
Finally, we  assemble the full result for the
boundary condition of the integral $I_{13}$ using
Eqs.~(\ref{eq30a},\ref{eq34a},\ref{eq36a}--\ref{eq41}).

We note that the computation of the other boundary integrals is performed following
similar steps; in fact, the manipulations of the hypergeometric functions
and the ``expansion by regions'' are similar  for all complicated master
integrals that one has to compute. We believe that the discussion of the
boundary condition of the integral $I_{13}$ provides enough insight into 
how to deal with them and we do not discuss other integrals
for that reason. We note that boundary conditions for all the
seventeen integrals $I_{1,..,17}$ are given in the Appendix.

\section{Final results}
\label{sec:finres}
With the boundary conditions known  and the system of linear equations
rationalized,  it is straightforward to integrate the equations in terms
of Goncharov polylogarithms and to match the result
of the integration to the  boundary conditions. 
This allows us to write the  function  $G_{ij}(z)$ in Eq.~\eqref{eq10}
as a linear combination of master integrals 
\be
G_{ij}(z) = \sum \limits R_{ij}^{(k)}(z) I_{k}(z),
\ee
where $R_{ij}^{(k)}(z)$ are the reduction coefficients
to master integrals.  
The integration over $z$ in Eq.~\eqref{eq10} is  straightforward --
we change variables $z \to t$ using Eq.~\eqref{eq15a} and
integrate from $t = (1-\sin\delta)/\cos\delta$, that corresponds
to $z=1$, to $t = \cos \delta$ that corresponds to $z=0$. 
The only subtlety is that the integration over $z$ diverges at
$z=0$. To overcome this problem, we write 
\be
\int \limits_{0}^{1}  {\rm d} z ~G_{ij}(z) =
\int \limits_{0}^{1}  {\rm d} z \left ( G_{ij}(z)  - \tilde G_{ij}(z)  \right )
+ \int \limits_{0}^{1} {\rm d} z ~\tilde G_{ij}(z),
\label{eq59new}
\ee
where $\tilde G_{ij}(z) \sim z^{-1-2\ep}$ describes the non-integrable behavior 
of the function $G_{ij}(z)$ at small $z$. This function can be extracted
from the computed boundary conditions for the master integrals and the small-$z$ expansion
of the reduction coefficients $R_{ij}^{(k)}(z)$.  Finally, since 
$\tilde G_{ij}(z) \sim z^{-1-2\ep}$, the last term
in Eq.~(\ref{eq59new}) can be trivially integrated over $z$ and, since the first
term is not singular at $z = 0$, the integrand can be expanded in $\ep$ and, after
changing variables from $z$ to  $t$, the integration over $t$ can be performed in a relatively
straightforward way. 

We note that after performing this  final $z$ (or, rather,  $t$)
integration, we obtain the result given by
a linear combination of Goncharov polylogarithms up to
weight four with
indices drawn from the following set
\be
\left \{a_\pm, b_\pm, e^{\pm i \delta}, \cos \delta,
\frac{1}{\cos \delta}, 0, -1, 1 \right \}, 
\ee
where $a_\pm, b_\pm$ are given in Eq.~\eqref{eq18a}. The arguments
of these Goncharov polylogarithms are either 
$(1-\sin \delta)/\cos \delta$ or  $\cos \delta$. 
The result of the $z$ integration (or rather $t$ integration) appears 
to be very large and complex.
However, it can be 
simplified using the (by now standard )
symbol techniques \cite{Goncharov:2010jf,Duhr:2011zq,Duhr:2012fh}.
Computing the symbol of the result, simplifying it and integrating
the result back, we arrive at the following expressions for the double-eikonal integrals 
\allowdisplaybreaks
\begin{align}
&\intSS_{ij}^{(gg)} = 
(2 E_{\rm max})^{-4\ep} \left[\frac{1}{8\pi^2}\frac{(4\pi)^\ep}{\Gamma(1-\ep)}\right]^2
\Bigg\{
\frac{1}{2\ep^4} + \frac{1}{\ep^3}\left[\frac{11}{12} - \lnss\right]
\notag\\
&
+\frac{1}{\ep^2}\left[2\Li_2(c^2) + \lnsk{2} -\frac{11}{6}\lnss + \frac{11}{3}\lnt
-\frac{\pi^2}{4}-\frac{16}{9}\right]
\notag\\
& 
+\frac{1}{\ep}\bigg[6\Li_3(s^2) + 2 \Li_3(c^2) + 
\left(2\lnss + \frac{11}{3}\right)\Li_2(c^2)
-\frac{2}{3}\lnsk{3}
\notag\\
&
\hphantom{+\frac{1}{\ep}\bigg[}
+\left(3\lncs + \frac{11}{6}\right)\lnsk{2}
-\left(\frac{22}{3}\lnt + \frac{\pi^2}{2} - \frac{32}{9}\right)\lnss
\notag\\
&
\hphantom{+\frac{1}{\ep}\bigg[}
-\frac{45}{4}\zeta_3-\frac{11}{3}\lntk{2} -\frac{11}{36}\pi^2
-\frac{137}{18}\lnt +\frac{217}{54}\bigg]
\notag\\
&
+4 \GPL{-1,0,0,1}{s^2} - 7 \GPL{0,1,0,1}{s^2}
+ \frac{22}{3}{\rm Ci}_3(2 \delta) + \frac{1}{3\tan(\delta)} {\rm Si}_2(2\delta)
\notag\\
&
+2 \Li_4(c^2)-14 \Li_4(s^2)+4\Li_4\left(\frac{1}{1+s^2}\right) 
-2 \Li_4\left(\frac{1-s^2}{1+s^2}\right) 
\notag\\
&
+ 2 \Li_4\left(\frac{s^2-1}{1+s^2}\right)
+\Li_4(1-s^4) 
+ \bigg[10 \lnss-4\lnopss
\notag\\
&
+\frac{11}{3}\bigg]\Li_3(c^2)
+ \left[14\lncs + 2\lnss + 4 \lnopss+\frac{22}{3}\right] \Li_3(s^2)
\label{eq:fingg}
\\
&
+ 4 \lncs \Li_3(-s^2)
+\frac{9}{2}\Li^2_{2}(c^2) - 4 \Li_2(c^2)\Li_2(-s^2)
+\bigg[7\lncs\lnss
\notag\\
&
-\lnsk{2}-\frac{5}{2}\pi^2+\frac{22}{3}\lnt-\frac{131}{18}\bigg]\Li_2(c^2)
+\bigg[\frac{2}{3}\pi^2- 4 \lncs\lnss\bigg]\times
\notag\\
&
 \Li_2(-s^2)
+\frac{\lnsk{4}}{3}+\frac{\lnopsk{4}}{6}
-\lnsk{3}\left[\frac{4}{3}\lncs+\frac{11}{9}\right]
\notag\\
&
+ \lnsk{2}
\bigg[7\lnck{2} + \frac{11}{3}\lncs + \frac{\pi^2}{3}+\frac{22}{3}\lnt-\frac{32}{9}\bigg]
-\frac{\pi^2}{6}\lnopsk{2}
\notag\\
&
 + \zeta_3\bigg[\frac{17}{2}\lnss - 11 \lncs + \frac{7}{2}
\lnopss-\frac{21}{2}\lnt - \frac{99}{4}\bigg]
+ \lnss\times\notag\\
&
\bigg[-\frac{7\pi^2}{2} \lncs + \frac{22}{3}\lntk{2}-\frac{11}{18}\pi^2
+\frac{137}{9}\lnt-\frac{208}{27}\bigg] - 12 \Li_4\left(\frac{1}{2}\right)
\notag\\
&
+\frac{143}{720}\pi^4 - \frac{\lntk{4}}{2} + \frac{\pi^2}{2}\lntk{2} 
-\frac{11}{6}\pi^2 \lnt + \frac{125}{216}\pi^2 + \frac{22}{9}\lntk{3}
\notag\\
&
+\frac{137}{18}\lntk{2} 
+ \frac{434}{27} \lnt - \frac{649}{81} + \mathcal O(\ep)
\Bigg\},\notag
\end{align}
and
\be
\begin{split}
&\intSS_{ij}^{(q\bar q)} = 
(2 E_{\rm max})^{-4\ep} \left[\frac{1}{8\pi^2}\frac{(4\pi)^\ep}{\Gamma(1-\ep)}\right]^2
\Bigg\{
-\frac{1}{3\ep^3} + \frac{1}{\ep^2}\bigg[\frac{2}{3}\lnss - \frac{4}{3}\lnt 
\\
&
+ \frac{13}{18}\bigg]
+\frac{1}{\ep}\bigg[
-\frac{4}{3}\Li_2(c^2)-\frac{2}{3}\lnsk{2} + \lnss
\left(\frac{8}{3}\lnt-\frac{13}{9}\right) + \frac{\pi^2}{9}
\\
&
+\frac{4}{3}\lntk{2}
+\frac{35}{9}\lnt-\frac{125}{54}\bigg]
-\frac{8}{3}{\rm Ci}_{3}(2\delta) - \frac{2}{3\tan(\delta)}{\rm Si}_{2}(2\delta)
-\frac{4}{3}\Li_3(c^2)
\\
&
-\frac{8}{3}\Li_3(s^2) + 
\Li_2(c^2)\left[\frac{29}{9}-\frac{8}{3}\lnt\right]
+\frac{4}{9}\lnsk{3}
+\lnsk{2}\bigg[-\frac{4}{3}\lncs
\\
&
-\frac{8}{3}\lnt+\frac{13}{9}\bigg]
+\lnss\bigg[-\frac{8}{3}\lntk{2}-\frac{70}{9}\lnt + \frac{2}{9}\pi^2 +
\frac{107}{27}\bigg]
+9\zeta_3 
\\
&
+ \frac{2\pi^2}{3}\lnt - \frac{8}{9}\lntk{3} - \frac{23}{108}\pi^2
-\frac{35}{9}\lntk{2} - \frac{223}{27}\lnt + \frac{601}{162}
+\mathcal O(\ep)
\Bigg\}.
\end{split}
\label{eq:finqqb} 
\ee
Here, $s=\sin \delta$, $c = \cos \delta$ and $\delta=\theta/2$
is half the angle between the three-momenta
of the hard radiators $i$ and $j$.  The Clausen functions are defined as
\begin{equation}
{\rm Ci}_n(z) = \frac{\left(\Li_n(e^{iz})+\Li_n(e^{-iz})\right)}{2},~~~~
{\rm Si}_n(z) = \frac{\left(\Li_n(e^{iz})-\Li_n(e^{-iz})\right)}{2i},
\end{equation}
and $\GPL{a_1,a_2,...,a_m}{x}$ are the standard
Goncharov polylogarithms. 
We have checked these analytic
results by computing the functions $\intSS_{ij}^{(gg,q\bar q)}$ numerically
for a few values of $\delta$ using the Mellin-Barnes representation
for the original eikonal integrals and the {\sf MB.m} routine for 
numerical integration \cite{Czakon:2005rk}. 
We found agreement within the expected numerical precision of the 
latter.\footnote{We are indebted to Ch. Wever for
  help with these checks.}

The results shown in Eqs.~(\ref{eq:fingg},\ref{eq:finqqb}) describe the integrals 
$\intSS_{ij}^{(gg,q\bar q)}$ as a function of the relative angle between the hard
emittors.  A useful special case corresponds to back-to-back
radiators; this kinematic situation 
is relevant for the description of color singlet production and decay. In  
the back-to-back limit $\delta = \theta/2 = \pi/2$
\be
\begin{split}
&\left.\intSS_{ij}^{(gg)}\right.|_{\delta\to\pi/2} = 
(2 E_{\rm max})^{-4\ep} \left[\frac{1}{8\pi^2}\frac{(4\pi)^\ep}{\Gamma(1-\ep)}\right]^2
\Bigg\{\frac{1}{2\ep^4} + \frac{11}{12\ep^3} 
+ \frac{1}{\ep^2}
\bigg[\frac{11}{3}\lnt
\\
&
- \frac{\pi^2}{4} -\frac{16}{9}\bigg]
+ \frac{1}{\ep}
\bigg[-\frac{21}{4}\zeta_3 - \frac{11}{3}\lntk{2} - \frac{137}{18}\lnt
-\frac{11}{36} \pi^2 + \frac{217}{54}\bigg]
\\
& 
-\frac{11}{80}\pi^4 -\frac{275}{12}\zeta_3 + \frac{22}{9}\lntk{3} 
- \frac{11}{6}\pi^2\lnt + \frac{125}{216}\pi^2 + \frac{137}{18} \lntk{2} 
\\
&
+ \frac{434}{27}\lnt -\frac{649}{81} + \mathcal O(\ep)\Bigg\},
\end{split}
\label{eq:b2bgg}
\ee
and
\be
\begin{split}
&\left.\intSS_{ij}^{(q\bar q)}\right.|_{\delta\to\pi/2} = 
(2 E_{\rm max})^{-4\ep} \left[\frac{1}{8\pi^2}\frac{(4\pi)^\ep}{\Gamma(1-\ep)}\right]^2
\Bigg\{-\frac{1}{3\ep^3} + \frac{1}{\ep^2}\bigg[\frac{13}{18}
\\
&
-\frac{4}{3}\lnt\bigg]
+\frac{1}{\ep}\bigg[\frac{\pi^2}{9}+\frac{4}{3}\lntk{2}+\frac{35}{9}\lnt-\frac{125}{54}
\bigg]
+\frac{25}{3}\zeta_3 + \frac{2\pi^2}{3}\lnt 
\\
&
- \frac{8}{9}\lntk{3} 
-\frac{35}{9}\lntk{2} -\frac{23}{108}\pi^2 - \frac{223}{27}\lnt + \frac{601}{162}
+\mathcal O(\ep) \Bigg\}.
\end{split}
\label{eq:b2bqq}
\ee
We have checked the back-to-back results Eqs.~(\ref{eq:b2bgg},\ref{eq:b2bqq})
against the numerical values used in Ref.~\cite{Caola:2017dug}, and found perfect
agreement. 

\section{Conclusions}
\label{sec:conclusions}
We  computed integrals  of
the double-soft eikonal functions over  phase-spaces of two soft
gluons or a soft $q \bar q$ pair  in the case when the
three-momenta of the  hard massless radiators  are at an arbitrary
angle to each other.  Within the framework of
a nested soft-collinear subtraction scheme \cite{Caola:2017dug}, 
our results will allow  for an analytic treatment of 
the double-soft contribution to NNLO QCD corrections to a  generic  process
with arbitrary number of hard massless color-charged particles.
Our  results for the integrated double-soft functions are compact; they  
are expressed 
in terms of ordinary and harmonic polylogarithms which ensures that they
can be evaluated numerically fast and efficiently. 
We look forward to the applications
of these results in NNLO QCD computations.

\vskip 1cm

\hspace*{-0.7cm}{\bf Acknowledgments} We would like to thank Ch. Wever for help
with some aspects of the computation. We are grateful to R. R\"ontsch
for useful discussions. 


\appendix

\section{Boundary conditions for master integrals}

We provide the boundary conditions for the required master integrals
in this Appendix.  We introduce the following notation 
\begin{align}
I_{\Gamma} = \frac{\GFP{3}{1-2\epsilon} \GF{1+\epsilon} }{\GFP{2}{1-\epsilon}\GF{1-3\epsilon}} = 1 + \epsilon^2 \frac{\pi^2}{6} - \epsilon^3  2 \zeta_3 - \epsilon^4 \frac{29\pi^4}{360} + \mathop{}\!\mathcal{O}\left(\eps^5\right), 
\end{align}
and write the results for master integrals as 
\be
I_i = N_\ep {\tilde I}_i,\;\;\;\; i = 1,...,17.
\ee
where $N_\ep$ is given in Eq.~\eqref{eq:nep}.
The integrals $\tilde I$ at small $z$ are given below. 
\begin{align*}
{\tilde I}_1 & = z^{1-2\epsilon}, \displaybreak[0] \\
{\tilde I}_2 & = z^{1-2\epsilon} \;  \frac{2 (1-2\epsilon)^2}{\epsilon}
 \times  \bigg\{ \frac{I_{\Gamma}z^{-\epsilon}}{1-3\epsilon}
- \frac1{1-2\epsilon}  \bigg\} + \ORD{z^2}, \displaybreak[0] \\
{\tilde I}_3 & = z^{1-2\epsilon} \;  4 (1-2\epsilon)^2  \times \\ 
&
\hphantom{=z^{1-2\ep}}\bigg\{ z^{-1-\epsilon}I_{\Gamma} \left[ \frac{-1}{2\epsilon}  
+ \frac{z (1-\epsilon)}{(1-3\epsilon)} \right] - \frac{2}{(1-2\epsilon)(1+\epsilon)}  \bigg\} + \ORD{z^2}, \displaybreak[0] \\
{\tilde I}_4 & = z^{-2\epsilon} \; \frac{2(1-2\epsilon)^2}{\epsilon^2} \times 
\bigg\{  1 -  z^{-\epsilon} I_{\Gamma} \HYPGF{1}{\epsilon}{1-\epsilon}{\cos ^2 \delta}  \bigg\} + \ORD{z}, \displaybreak[0] \\
{\tilde I}_5 & = z^{-2\epsilon} \; \frac{-6 ( 1- 2 \epsilon)^2}{\epsilon} z^{-1-\epsilon} I_{\Gamma} \times \HYPGF{1}{1+\epsilon}{1-\epsilon}{\cos^2 \delta }  + \ORD{z^0},  \displaybreak[0] \\
{\tilde I}_6 & =  z^{-1-2\epsilon} \; \frac{(1-2\epsilon)^2}{\epsilon^2} \times \Bigg\{-3 z^{-\epsilon} I_{\Gamma} \HYPGF{1}{1+\epsilon}{1-\epsilon}{\cos^2\delta} \Bigg\} \\
& + 2 \tilde{I}_{18} + \ORD{z^0}, \displaybreak[0] \\
{\tilde I}_7 & = z^{-1-2\epsilon} \; \frac{(1-2\epsilon)^2}{\epsilon^2} \times \Bigg\{ 2 + 2 \ z^{1-\epsilon} I_{\Gamma} \HYPGF{1}{\epsilon}{1-\epsilon}{\cos^2\delta}  \Bigg\} \\*
& + \frac{2(1-2\ep)^2(1+4\ep)}{z\ep^2(1+2\ep)} \tilde{I}_1 - \frac{(1+3\ep)z}{\ep} \tilde{I}_{18} + \frac{(1+\ep)z\sin^2 \delta}{\ep(1+2\ep)}\tilde{I}_{19} + \ORD{z^0}, \displaybreak[0] \\
{\tilde I}_8 & = z^{1-2\epsilon} \; \frac{(1-2\epsilon)}{\epsilon} \frac1{\sin ^ 2 \delta} \times \bigg\{ 2 z^{-\epsilon}  \frac{ (1-2\epsilon)}{(1-3\epsilon)} I_{\Gamma}  
\\
& -  4 \left[ \sin ^2 \delta \right]^{-\epsilon} \HYPGF{-\epsilon}{-\epsilon}{1-\epsilon}{\cos ^2 \delta} \bigg\} + \ORD{z^2}, \displaybreak[0] \\
{\tilde I}_9 & = z^{1-2\epsilon}  \; 8(1-2\epsilon)^2  \times \bigg\{ - \frac{(1+2\epsilon)}{(1+\epsilon)(2+\epsilon)(1-2\epsilon)} \HYPGF{1}{2}{3+\epsilon}{\sin^2 \delta} \\*
& - \frac{z^{-1-\epsilon}I_{\Gamma}}{\left[ \sin \delta \right]^2} \left[\frac1{4\epsilon} + \frac{z(1+\epsilon)}{2(1-3\epsilon)} \right]  +  \frac{z^{-\epsilon}I_{\Gamma}}{\left[ \sin \delta \right]^4} \frac{(1+\epsilon)}{2\epsilon(1-3\epsilon)} \\*
& - \frac{\left[\cos\delta\right]^{2\epsilon}}{\left[\sin\delta\right]^{4+2\epsilon}} \GF{1-\epsilon}\GF{1+\epsilon} \frac{(1+2\epsilon)}{\epsilon(1-2\epsilon)} \bigg\} + \ORD{z^2}, \displaybreak[0] \\
{\tilde I}_{10} & = z^{-2\epsilon} \;  \frac{(1-2\epsilon)^2}{\epsilon^2} \frac1{\sin^2 \delta} \times \ \bigg\{ 4 \left[\sin^2 \delta\right]^{-\epsilon}  \HYPGF{-\epsilon}{-\epsilon}{1-\epsilon}{\cos^2\delta} 
\\
&
-  z^{-\epsilon} I_{\Gamma}  \bigg\} + \ORD{z}, \displaybreak[0] \\
{\tilde I}_{11} & = \ORD{z}, \displaybreak[0] \\
{\tilde I}_{12} & = \ORD{z}, \displaybreak[0] \\
{\tilde I}_{13} & = z^{1-2\epsilon} \; 16(1-2\epsilon)^2 \times \bigg\{ - \frac{(1+2\epsilon)}{(1+\epsilon)(2+\epsilon)(1-2\epsilon)} \HYPGF{1}{2}{3+\epsilon}{\sin^2 \delta} \\*
& - \frac{z^{-1-\epsilon}I_{\Gamma}}{\left[ \sin \delta \right]^2} \left[\frac1{4\epsilon} - \frac{z(1+\epsilon)(1-2\epsilon)}{4\epsilon(1-3\epsilon)} \right]  +  \frac{z^{-\epsilon}I_{\Gamma}}{\left[ \sin \delta \right]^4} \frac1{\epsilon(1-3\epsilon)} \\*
& - \frac{\left[\cos\delta\right]^{2\epsilon}}{\left[\sin\delta\right]^{4+2\epsilon}} \GF{1-\epsilon}\GF{1+\epsilon} \frac{(1+2\epsilon)}{\epsilon(1-2\epsilon)} \bigg\} + \ORD{z^2}, \displaybreak[0] \\
{\tilde I}_{14} & =  z^{1-3\epsilon} \;  4 (1-2\epsilon)^2 \ \times  \ \frac{I_{\Gamma} }{\epsilon(1-3\epsilon)} \frac1{\sin^2\delta} \bigg\{ 2\epsilon - \sin^2\delta - \epsilon \sin^2\delta \bigg\} + \ORD{z^2}, \displaybreak[0] \\
{\tilde I}_{15} & = z^{1-2\epsilon} \; 4(1-2\epsilon)^2 \times \bigg\{ - \frac{2(1+2\epsilon)}{(1+\epsilon)(2+\epsilon)(1-2\epsilon)} \HYPGF{1}{2}{3+\epsilon}{\sin^2 \delta} \\*
& + \frac{z^{-1-\epsilon}I_{\Gamma}}{\left[ \sin \delta \right]^2} \left[\frac1{2\epsilon^2} - \frac{z(1+\epsilon)(1-\epsilon+2\epsilon^2)}{2\epsilon^2(1-3\epsilon)(1-\epsilon)} \right]  +  \frac{z^{-\epsilon}I_{\Gamma}}{\left[ \sin \delta \right]^4} \frac{(3-\epsilon)}{\epsilon(1-\epsilon )(1-3\epsilon)} \\*
& - \frac{\left[\cos\delta\right]^{2\epsilon}}{\left[\sin\delta\right]^{4+2\epsilon}} \GF{1-\epsilon}\GF{1+\epsilon} \frac{2(1+2\epsilon)}{\epsilon(1-2\epsilon)} \bigg\} + \ORD{z^2}, \displaybreak[0] \\
{\tilde I}_{16} & = z^{-2\epsilon} \; \frac{8(1-2\epsilon)^2}{\epsilon^2} \times \bigg\{ \frac{\epsilon}{(1+\epsilon)} \HYPGF{1}{1}{2+\epsilon}{\sin^2 \delta} \\*
& +  \left[ \sin ^2 \delta \right]^{-1-\epsilon}  \GF{1+\epsilon} \GF{1-\epsilon} \;   \left[ \cos^2 \delta \right]^\epsilon  \\*
& - \frac1{4} \left[\sin^2\delta \right]^{-1} z^{-\epsilon} I_{\Gamma} \left[1 + 2 \; \HYPGF{1}{\epsilon}{1-\epsilon}{\cos^2\delta} \right]   \bigg\} + \ORD{z}, \displaybreak[0] \\
{\tilde I}_{17} & = z^{-2\epsilon} \; 8(1-2\epsilon)^2 \times  \Bigg\{  \frac{(1+2\epsilon)}{\epsilon(1+\epsilon)(2+\epsilon)} \HYPGF{1}{2}{3+\epsilon}{\sin^2 \delta} 
+\frac{z^{-1-\epsilon}I_{\Gamma}}{\left[ \sin \delta \right]^2}\times
\\*
& \bigg[\frac{3}{4\epsilon^2} 
\times \HYPGF{1}{1+\epsilon}{1-\epsilon}{\cos^2\delta} + \frac{z(1+\epsilon)}{2\epsilon(1-\epsilon)} \times 
\HYPGF{1}{\epsilon}{1-\epsilon}{\cos^2\delta} \bigg] 
\\*
&
 -  \frac{z^{-\epsilon}I_{\Gamma}}{\left[ \sin \delta \right]^4} \left[ \frac1{2\epsilon^2} + \frac{(1+\epsilon)}{2\epsilon^2(1-\epsilon)}  \times \HYPGF{1}{\epsilon}{1-\epsilon}{\cos^2\delta} \right] \\*
& + \frac{\left[\cos\delta\right]^{2\epsilon}}{\left[\sin\delta\right]^{4+2\epsilon}} \GF{1-\epsilon}\GF{1+\epsilon} \frac{(1+2\epsilon)}{\epsilon^2} \bigg\} + \ORD{z}.
\end{align*}
We note that the hard contributions for master integrals $\tilde{I}_6$ and $\tilde{I}_7$
can be written in terms of the master integrals $\tilde{I}_{18}$ and
$\tilde{I}_{19}$, see Eqs.~(\ref{eq18aa}). This provides an easy way to compute these
integrals to the
required order in the $\epsilon$-expansion.

\end{document}